\documentclass[aps,prb,twocolumn,showpacs,superscriptaddress]{revtex4}
\usepackage{graphicx,amssymb,amsfonts,amsmath}

\begin{document}

\title{A random phase approximation study of one-dimensional fermions after a quantum quench}
\author{Jarrett Lancaster}
\affiliation{Department of Physics, New York University,
4 Washington Place, New York, New York 10003, USA}
\author{Thierry Giamarchi}
\affiliation{DPMC-MaNEP, University of Geneva, 24 Quai Ernest-Ansermet, CH-1211
 Geneva, Switzerland}
\author{Aditi Mitra}
\affiliation{Department of Physics, New York University,
4 Washington Place, New York, New York 10003, USA}
\date{\today}

\begin{abstract}
The effect of interactions on a system of fermions that are in a non-equilibrium steady state due
to a quantum quench is studied employing the random-phase-approximation (RPA). As a result of the
quench, the distribution function of the fermions is highly broadened. This gives rise
to an enhanced particle-hole spectrum and over-damped collective modes for
attractive interactions between fermions. On the other hand, for repulsive interactions, an undamped
mode above the particle-hole continuum survives. The sensitivity of the result on the
nature of the non-equilibrium steady state is explored by also considering a quench that
produces a current carrying steady-state.
\end{abstract}

\maketitle

\section{Introduction}

Recent remarkable experiments~\cite{weiss} with cold atoms have motivated an explosion
of theoretical interest in the area of non-equilibrium quantum dynamics with a
focus on addressing fundamental questions about thermalization,
chaos and integrability, issues that are very relevant to these experimental systems.~\cite{polkovrev10}
Without many general results on generic non-equilibrium phenomena, the analysis of specific,
tractable models is a common way to make progress. One hopes clues gathered from these specific
systems will lead to more general predictions.

One-dimensional (1D) systems are where much of the theoretical work has taken place since a wide
array of tools is available for investigating dynamics. An interesting class of these systems is
integrable models, where conserved quantities tightly constrain the time evolution. While a consensus lacks
on a rigorous definition of quantum integrability,~\cite{caux2011} progress has been made using
many quantum models satisfying classical notions of integrability. Fruitful studies have investigated dynamics
of Bethe-ansatz solvable models,~\cite{mosselcaux2010} but the simplest integrable models are the
quadratic ones. These effectively non-interacting theories, including those considered in this paper,
allow for exact analytical treatment of the non-trivial
dynamics.~\cite{jllam1,antal2,caz1,iucci,caz2,kennes2010}
In 1D, efficient numerical studies are also possible with the time-dependent density matrix
renormalization group (tDMRG),\cite{tdmrg,tdmrg2} and exact diagonalization studies
of finite systems.~\cite{rigol2009,santosrigol2010}

Some of the analytical and numerical studies have revealed that 1D
systems after a quantum quench often reach athermal steady states which can be characterized by
a generalized Gibbs ensemble (GGE) constructed from
identifying the conserved quantities of the system.~\cite{Rigol07,caz1,iucci,cardy,kennes2010}
There are also many counter-examples where such a description
fails, as not all physical quantities can be described using the
GGE.~\cite{jllam1,iucci,Barthel08,Gangardt08,Silva10,Silva11}

One important question concerns the stability of these athermal steady states generated
after a quantum quench
to other perturbations such as non-trivial interactions that introduce mode-coupling
and/or the breaking of integrability.
Precisely this question was addressed recently in Ref~\onlinecite{mitra11b}. In particular
an initial interaction quench in a Luttinger liquid gives rise
to an athermal steady state characterized by new power-law exponents,~\cite{caz1,iucci,mitra11b}
which can also be captured using a GGE.
The effect of mode-coupling arising due to a periodic potential on this
non-equilibrium state
was studied in Ref.~\onlinecite{mitra11b} using perturbative renormalization group. The
analysis revealed that infinitesimally small perturbations can generate
not only an effective-temperature but also a dissipation or a finite lifetime of the bosonic
modes. While the appearance of an effective-temperature, although highly non-trivial in itself,
can be rationalized on the grounds that a
system after a quench is in a highly excited state,
and that interactions between particles will somehow cause the system to ``thermalize'',
the appearance of dissipation is an unexpected and non-trivial result. Thus one of the motivations
of the current paper is to identify other physical situations where this dissipation might appear,
and to try to investigate the physical mechanisms that could be behind it.
Due to the close parallels between interacting bosons and fermions in 1D, a natural candidate
for analyzing this
question is a one-dimensional system of free fermions that is in a non-equilibrium steady state after a
quench. We analyze the effect of weak interactions on this system by employing the
random-phase-approximation (RPA).

In equilibrium, 1D systems are the ideal playground for invoking the RPA.
It is an exact low-energy treatment of weak interactions in 1D.\cite{larkin,everts,dover,Giamarchibook}
In particular, by applying RPA to a 1D system of electrons~\cite{dassarma2,dassarma} one recovers
the standard bosonization of the model, described by a Luttinger liquid.\cite{Giamarchibook} 
Note that this direct equivalence only holds
for the long wavelength properties, while other excitations require more sophisticated methods such as bosonization.
While the accuracy of RPA in 1D is known in equilibrium, its applicability out of
equilibrium is not guaranteed. In the type of quench problem considered here, the initial state
has nonzero overlap with excited eigenstates of the Hamiltonian generating time evolution. It is far from
certain that a low-energy description captures all the important physics. While this caveat leads to intriguing,
unanswered questions, we will use in this paper RPA as an approximation scheme and will not address the deeper question of
its potential breakdown out of equilibrium.

In this paper, we thus apply the RPA to a non-equilibrium state in the $XXZ$ spin-chain. This
state is prepared as follows. The system is initially in the ground state of an
exactly solvable Hamiltonian $H_i$. We choose two different models for $H_i$,
one corresponding to the transverse-field Ising model with the
magnetic field tuned to the critical value where the spectrum is gapless, and the
second is the same as above but with an additional Dzyaloshinskii-Moriya interaction added.
A quantum quench is then performed by switching
off the field and changing the exchange anisotropy so that the time evolution is due to the $XX$ model.
Since this model is described by free fermions, at long times
after the quench, the system reaches an athermal steady state
characterized by a GGE. For $H_i$ that has Dzyaloshinskii-Moriya interactions, the steady state
is qualitatively different in that it carries a net current.
We then ask how these athermal steady states are affected by weak Ising
interactions of the $XXZ$ chain ($J^z\sum_j {\hat S}^z_j{\hat S}^z_{j+1}$)
which are assumed to have been switched on very slowly.
The effects of the Ising interactions are treated using RPA.

We demonstrate the existence of a single undamped collective mode
for repulsive interactions ($J_z > 0$) which is qualitatively similar
to the predictions of the RPA in equilibrium, but with some quantitative changes
to the mode-velocity. On the other hand for attractive interactions ($J_z <0$), no undamped modes
are found for both the athermal states that have been studied.
This is because immediately after the quench the distribution function of
the fermions is highly broadened, thus creating an
enlarged particle-hole continuum.
As a consequence for attractive interactions either no solutions are found, or only damped solutions that
lie in the particle-hole continuum are found.
Further, if the current in the athermal steady state is larger than a critical value, then even the
undamped mode for repulsive interactions vanishes in the long-wavelength limit.

These results are consistent with the ones obtained in Ref~\onlinecite{mitra11b} where it was
found that as a result of a quantum quench and mode-coupling, a Luttinger
liquid description is replaced by a low energy effective theory of thermal bosons with a finite lifetime.
Indeed, in equilibrium, the undamped collective modes obtained from RPA can also be described as a Luttinger
liquid.~\cite{Giamarchibook} The RPA analysis shows that for attractive interactions the collective
modes lie in the particle-hole continuum and are therefore overdamped.
The analysis of the present paper thus allows one to
interpret the generation of the friction that was put in evidence by the RG analysis of
Ref~\onlinecite{mitra11b} as
due to a generalization to an out of equilibrium case of Landau damping.

The paper is organized as follows. In section~\ref{model} the models that will be studied and the
notations and conventions are defined. In section~\ref{quench1} the RPA analysis where
the quench is from the gapless phase of the transverse-field Ising model to the $XX$ model is considered.
In section~\ref{quench2} RPA involving fermions in a current carrying steady state is presented,
and in section~\ref{conclu} we summarize our results.

\section{Model} \label{model}

Below we describe the two
different quenches which lead to non-equilibrium steady states without (sub-section~\ref{IIa}) and
with (sub-section~\ref{IIb}) currents.

\subsection{Quench from ground state} \label{IIa}

The
$XY\,$ spin chain in a magnetic field is defined as
\begin{eqnarray}
\hat{H}_{i} & = &-J\sum_{j}\left[(1+\gamma)\hat{S}^{x}_{j}\hat{S}_{j+1}^{x}
+(1-\gamma)\hat{S}_{j}^{y}\hat{S}_{j+1}^{y}\right]\nonumber\\
& + & h\sum_{j}\hat{S}_{j}^{z}.
\end{eqnarray}
where $\gamma=1$ corresponds to the transverse-field Ising model.
The $XY$ model has been extensively studied,\cite{lsm,barouch3,taylor} and its equilibrium properties are
well understood. It is also a popular model\cite{cardy,levitov,prosen} for studying non-equilibrium situations
due to its simple mapping to free fermions. Writing this Hamiltonian in terms of Jordan-Wigner
fermions,\cite{lsm}
\begin{eqnarray}
\hat{S}_{j}^{+} & = & c^{\dagger}_{j}\exp\left[i\pi\sum_{n<j}c_{n}^{\dagger}c_{n}\right],\\
\hat{S}_{j}^{-} & = & \exp\left[-i\pi\sum_{n<j}c_{n}^{\dagger}c_{n}\right]c_{j},\\
\hat{S}^{z}_{j} & = & c_{j}^{\dagger}c_{j} - \frac{1}{2}.
\end{eqnarray}
we obtain
\begin{eqnarray}
\hat{H}_{i} & = & -\frac{J}{2}\sum_{j}\left[c_{j}^{\dagger}c_{j+1}+c_{j+1}^{\dagger}c_{j}+\gamma c_{j}^{\dagger}c_{j+1}^{\dagger}+\gamma c_{j+1}c_{j}\right]\nonumber\\
& + & h\sum_{j}c_{j}^{\dagger}c_{j}.
\end{eqnarray}
This is diagonalized by a Bogoliubov rotation\cite{barouch3}
\begin{equation}
\hat{H}_{i}  =  \sum_{k}\epsilon_{k}^{i}\eta_{k}^{\dagger}\eta_{k},
\end{equation}
where
\begin{eqnarray}
\epsilon_{k}^{i} = -J\mbox{sign}\left(\cos k -\frac{h}{J}\right)\sqrt{\left(\cos k -\frac{h}{J}\right)^{2}
+ \gamma^{2}\sin^{2}k} \label{disp}
\end{eqnarray}
and
\begin{equation}
\left(\begin{array}{c} c_{k}\\ c_{-k}^{\dagger}\end{array}\right) = \left(\begin{array}{cc} \cos\frac{\theta_{k}}{2} & -i\sin\frac{\theta_{k}}{2}\\ -i\sin\frac{\theta_{k}}{2} & \cos\frac{\theta_{k}}{2}\end{array}\right)\left(\begin{array}{c} \eta_{k}\\ \eta_{-k}^{\dagger}\end{array}\right).
\end{equation}
with
\begin{eqnarray}
\cos\theta_{k} & = &  \frac{|\cos k - (h/J)|}{\sqrt{\left(\cos k -\frac{h}{J}\right)^{2} + \gamma^{2}\sin^{2}k}},\\
\sin\theta_{k} & = & \frac{\mbox{sign}(\cos k - (h/J))\gamma \sin k}{\sqrt{\left(\cos k -\frac{h}{J}\right)^{2} + \gamma^{2}\sin^{2}k}}.
\end{eqnarray}
Here $c_{j} = \frac{1}{\sqrt{N}}\sum_{k}e^{ikj}c_{k}$. The ground state is obtained by occupying all modes with negative energy. We will be interested in the special case of $\gamma = h/J = 1$, where the system is critical.\cite{sachdev} The ground state is defined by $\eta_{k}\left|\Phi_{0}\right\rangle = 0\,$ for all $k$, as $\epsilon_{k}^{i}=2J\left|\sin\frac{k}{2}\right|\,$ is always non-negative.

Given this initial state, we perform the quench by suddenly switching off the anisotropy $\gamma\,$
and magnetic field $h$. The subsequent time evolution is due to the $XX$-Hamiltonian,
\begin{eqnarray}
\hat{H}_{XX} & = & -J\sum_{j}\left[\hat{S}^{x}_{j}\hat{S}^{x}_{j+1}+\hat{S}^{y}_{j}\hat{S}^{y}_{j+1}\right]\\
& = & \sum_{k}\epsilon_{k}c_{k}^{\dagger}c_{k},
\end{eqnarray}
where $\epsilon_{k} = -J\cos k\,$ and the $c_{k}\,$ are the momentum-space Jordan-Wigner fermions defined above. At long times after the quench, the system approaches a diagonal ensemble. To see this, note that immediately after the quench, the following quantities are fixed by the initial state
\begin{eqnarray}
\left\langle c_{k}^{\dagger}c_{k}\right\rangle_{0} & = & \cos^{2}\frac{\theta_{k}}{2}\left\langle\eta_{k}^{\dagger}\eta_{k}\right\rangle + \sin^{2}\frac{\theta_{k}}{2}\left\langle \eta_{-k}\eta_{-k}^{\dagger}\right\rangle\\
\left\langle c_{k}c_{-k}\right\rangle_{0} & = & i\sin\frac{\theta_{k}}{2}\cos\frac{\theta_{k}}{2}\left[\left\langle \eta_{k}\eta_{k}^{\dagger}\right\rangle - \left\langle \eta_{-k}^{\dagger}\eta_{-k}\right\rangle\right].
\end{eqnarray}
Time evolution of the $c$-operators takes the simple form $c_{k}(t) = e^{-i\epsilon_{k}t}c_{k}$. One finds
\begin{eqnarray}
\left\langle c_{k}^{\dagger}(t)c_{k}(t)\right\rangle & = & \left\langle
c_{k}^{\dagger}c_{k}\right\rangle_{0},\\
\left\langle c_{k}(t)c_{-k}(t)\right\rangle & = & e^{-2i\epsilon_{k}t}\left\langle c_{k}c_{-k}\right\rangle_{0}.
\label{oscill}
\end{eqnarray}
When averaged over long times,
\begin{eqnarray}
\frac{1}{t_{b}-t_{a}}\int_{t_{a}}^{t_{b}}dt\left\langle c_{k}(t)c_{-k}(t)\right\rangle \rightarrow 0
\end{eqnarray}
for $(t_{b}-t_{a})\rightarrow \infty$, due to the rapidly oscillating exponential. Thus we obtain
a diagonal ensemble in the long-time limit with a highly broadened momentum distribution given
by
\begin{equation}
\left\langle c_{k}^{\dagger}c_{k}\right\rangle_{0} = \frac{1}{2}\left(1-\left|\sin\frac{k}{2}\right|\right).
\end{equation}
Note that, as we discuss at the end of this section, such an approximation is not necessary and one
can retain the oscillating modes. It however
considerably simplifies the expressions to explicitly eliminate them.
By equating the above non-equilibrium distribution function to a Fermi function,
one may define a (momentum dependent) effective temperature~\cite{Essler11}
\begin{equation}
T_{\mbox{\scriptsize eff}}^{c_{k}} = -J\cos k \left(\ln\left[\frac{1 +|\sin\frac{k}{2}|}
{1-|\sin\frac{k}{2}|}\right]\right)^{-1},
\label{eq:teffc}
\end{equation}
where one notes $T_{\mbox{\scriptsize eff}}^{c_{k}}<0$ for $k<\pi/2$. As we
shall see later, since the system is
out of equilibrium, this temperature is not universal, but depends on
the quantity being studied.

Once the above steady state has been reached,
we consider the effect of nearest-neighbor Ising interactions in the $XXZ\,$
model
\begin{equation}
\hat{H}_{f}  =  \hat{H}_{XX} + J^{z}\sum_{j}\hat{S}_{j}^{z}\hat{S}_{j+1}^{z},
\end{equation}
where we assume that $J^z$ was switched on very slowly, so that in the absence of
a quench, the fermions will evolve into the ground state of the $XXZ$ chain.
The effects of this interaction term will be treated within the RPA.

The basic fermionic Green's functions defined by
\begin{eqnarray}
G^R_f(k;t,t^{\prime}) & = & -i\theta(t-t^{\prime})\langle \{c_k(t),c_k^{\dagger}(t^{\prime})\}\rangle,\\
G^K_f(k;t,t^{\prime}) & = & -i\langle \left[c_k(t),c_k^{\dagger}(t^{\prime})\right]\rangle,
\end{eqnarray}
are found to be
\begin{eqnarray}
G^R_f(k;t,t^{\prime}) & = & -i\theta(t-t^{\prime})e^{-i\epsilon_k(t-t^{\prime})},\label{Gr}\\
G^K_f(k;t,t^{\prime}) & = & -ie^{-i\epsilon_k(t-t^{\prime})}\left|\sin\frac{k}{2}\right|\mbox{sign}(\epsilon_k^{i}).
\end{eqnarray}

Within the RPA, the particle-hole bubbles are\cite{kamenev}
\begin{eqnarray}
\Pi^R(1,2) & = & \frac{-i}{2}\left[G^R_f(1,2) G^K_f(2,1)\right.\nonumber\\
& + & \left. G^K_f(1,2) G^A_f(2,1)\right],\\
\Pi^K(1,2) & = & \frac{-i}{2}\left[G^K_f(1,2) G^K_f(2,1) + G^R_f(1,2) G^A_f(2,1)\right.\nonumber\\
& + & \left.G^A_f(1,2) G^R_f(2,1)\right].
\end{eqnarray}
which in frequency-momentum space are given by,
\begin{eqnarray}
\Pi^{R}(q,\omega) & = & \frac{-i}{2}\int\frac{dk}{2\pi}\frac{d\Omega}{2\pi}
\left[G^{R}_f(k+q,\omega+\Omega)G^{K}_f(k,\Omega)\right.\nonumber\\
& + & \left.G^{K}_f(k+q,\omega+\Omega)G^{A}_f(k,\Omega)\right]\\
\Pi^{K}(q,\omega) & = & \frac{-i}{2}\int\frac{dk}{2\pi}\frac{d\Omega}{2\pi}
\left[G^{K}_f(k+q,\omega+\Omega)G^{K}_f(k,\Omega)\right.\nonumber\\
& + &  G^{R}_f(k+q,\omega+\Omega)G^{A}_f(k,\Omega) \nonumber\\
& + & \left.G^{A}_f(k+q,\omega+\Omega)G^{R}_f(k,\Omega)\right].
\end{eqnarray}
The collective mode dispersion is defined by the roots of the complex dielectric function,\cite{dassarma}
\begin{equation}
\epsilon_{\mbox{\scriptsize RPA}}(q,\omega_{q}) =1-V_{q}\Pi^{R}(q,\omega_{q}) = 0,
\end{equation}
where we neglect the $q$-dependence of $V_q$ and take
it to be $V_{q}= J^{z}\equiv V_{0}$.
The RPA analysis is given in Section~\ref{quench1}.

As mentioned above, it is not critical to work with the
diagonal ensemble. If we had retained the full time-dependence in Eq.~(\ref{oscill}), the integration
over the internal $k$ variable in the evaluation of the RPA bubbles would result in
terms that decay with time. Since we are ultimately interested in the long-time limit, these
contributions are not important for us.

\subsection{Quench resulting in a current-carrying state}\label{IIb}

We will also be interested in how the collective dynamics change when the athermal steady state is
characterized by a net current. This is generated by adding a
Dzyaloshinskii-Moriya interaction term\cite{dm} to the $XY$ model,
$\hat{H}_{i}\rightarrow \hat{H}_{i} + \hat{H}_{DM}$, where
\begin{equation}
\hat{H}_{DM} = -\lambda\sum_{j}\left[\hat{S}_{j}^{y}\hat{S}_{j+1}^{x} - \hat{S}_{j}^{x}\hat{S}_{j+1}^{y}\right].
\end{equation}
This new Hamiltonian is diagonalized by the same Bogoliubov
rotation\cite{derzhko,siskens} that diagonalizes the pure $XY$-model.
In the isotropic case ($XX\,$ chain), this can be interpreted as a spatially dependent,
physical rotation of the spins.\cite{perk1976,antal1} For the more general anisotropic chain,
the spectrum is similarly modified\cite{derzhko,das2011}
\begin{eqnarray}
\hat{H}_{i}(\lambda) & = & \sum_{k}\epsilon^{\prime}_{k}\eta_{k}^{\dagger}\eta_{k},\\
\epsilon^{\prime}_{k} & = & \epsilon_{k}^{i} -\lambda\sin k.
\end{eqnarray}
with $\epsilon_k^i$ given in Eq.~(\ref{disp}).
$\lambda$ has the effect of raising the energies of states with $k<0\,$ while lowering the energies of
modes with positive $k$. The occupation number is now nonzero for $\lambda >J$,
and the $\eta$-fermion occupation is
$n_{k} \equiv \left\langle \eta_{k}^{\dagger}\eta_{k}\right\rangle = \theta(k)\theta(k_{0}-k)$, with
\begin{equation}
k_{0} = 2\cos^{-1}\frac{J}{\lambda}.
\end{equation}

We will see in section~\ref{quench2} that the presence of this non-zero ``Fermi momentum''
will give rise to multiple damped modes within the particle-hole continuum
that are not present for the zero-current steady-state. Furthermore, above a certain
critical filling factor the single undamped collective mode will cease to exist.

The asymmetry in momentum space drives a current in the modified ground state given by
\begin{equation}
\left\langle j_{n}\right\rangle = J\mbox{Im}\left\langle\hat{S}^{+}_{n+1}\hat{S}_{n}^{-}\right\rangle = \left\{\begin{array}{cc} 0 & (\lambda < J)\\
\frac{J}{\pi}\left(1-\left(\frac{J}{\lambda}\right)^{2}\right) & (\lambda > J)\end{array}\right.\label{I}
\end{equation}
It should be noted that this operator can be interpreted as the current operator only within the $XX$-model where the total magnetization commutes with the Hamiltonian.

\begin{figure}
\includegraphics[totalheight=5cm,]{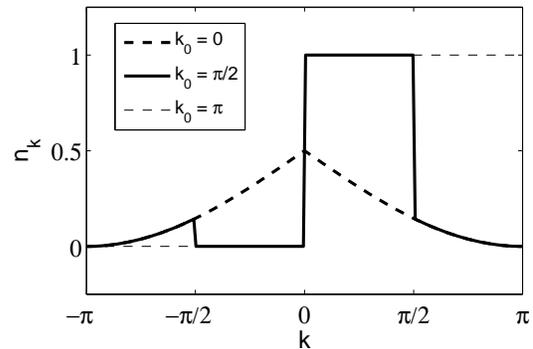}
\caption{Initial $c$-fermion distributions for (a) no current ($k_0=0$), (b) non-zero current
($k_0=\pi/2$), and (c) maximum current ($k_0=\pi$).
Note the sharp discontinuity for the case of non-zero current.}
\label{dist}
\end{figure}

Performing a quench where $\lambda$, $h$ and $\gamma$ are switched off allows this state
to evolve under the $XX\,$ model, obtaining a non-equilibrium momentum distribution
\begin{eqnarray}
\left\langle c_{k}^{\dagger}c_{k}\right\rangle_{0} & = & \frac{1}{2}\left(1-\left|\sin\frac{k}{2}\right|\right)\nonumber\\
& + & \frac{\theta(k)\theta(k_{0}-k)}{2}\left(1+\left|\sin\frac{k}{2}\right|\right)\nonumber\\
& - & \frac{\theta(-k)\theta(k_{0}+k)}{2}\left(1-\left|\sin\frac{k}{2}\right|\right)\label{dist2}.
\end{eqnarray}
and a current given by Eq.~(\ref{I}).
The distribution function for
several different current strengths is shown
in Fig.~\ref{dist}.

After the decay of transients $\left\langle c_{k}c_{-k}\right\rangle$,
$\left\langle c^{\dagger}_{k}c^{\dagger}_{-k}\right\rangle$, we investigate the collective modes by employing the
RPA analysis outlined in sub-section~\ref{IIa}.
The RPA requires knowledge of the single-particle Green's functions.
The presence of a current does not affect the
retarded Green's function (Eq.~(\ref{Gr})), but modifies the Keldysh Green's function as follows
\begin{eqnarray}
G^{K}_{f}(k;t,t') & = & -ie^{-i\epsilon_{k}(t-t')}\left[\left|\sin\frac{k}{2}\right|(1-n_{k}-n_{-k})\right.\nonumber\\
& - & \left.(n_{k}-n_{-k})\right],
\end{eqnarray}
where $n_{k} = \theta(k)\theta(k_{0}-k)\,$ is the occupation number of the $\eta$-fermions in the initial state.

\section{RPA for quench from ground state} \label{quench1}

In this section, we investigate the effect of interactions on the athermal steady state
(sub-section~\ref{IIa}) obtained from quenching from the ground state of
the transverse-field Ising model.
The RPA particle-hole bubbles are

\begin{eqnarray}
\Pi^{R}(q,\omega) & = &
- \frac{1}{2}\int \frac{dk}{2\pi}\left[\frac{\cos\theta_{k}(1-2n_{k})}{(\omega+i\delta)
+\epsilon_{k}-\epsilon_{k+q}}\right.\nonumber\\
& - & \left.\frac{\cos\theta_{k+q}(1-2n_{k+q})}{(\omega+i\delta) +\epsilon_{k}-\epsilon_{k+q}}\right]
\label{pir1}.
\end{eqnarray}
\begin{eqnarray}
\Pi^{K}(q,\omega)  =  \frac{i}{2}(2\pi)\int\frac{dk}{2\pi}\delta(\omega + \epsilon_{k}-\epsilon_{k+q})\nonumber\\
\times [\cos\theta_{k}\cos\theta_{k+q}(1-2n_{k})(1-2n_{k+q})-1].
\end{eqnarray}
For the case of interest, $\gamma = h/J = 1$, we have $\cos\theta_{k} = \left|\sin\frac{k}{2}\right|$,
and the distribution of the $\eta$ fermions $n_k=0$.

There are some basic symmetries of the
polarization bubbles that are worth mentioning. Firstly $\Pi^{R,K}(q,\omega)= \Pi^{R,K}(-q,\omega)$,
while ${\rm Re}[\Pi^R](q,-\omega) = {\rm Re}[\Pi^R](q,\omega),\Pi^K(q,-\omega)
= \Pi^K(q,\omega)$ and
${\rm Im}[\Pi^R(q,-\omega)] = -{\rm Im}[\Pi^R](q,\omega)$. Therefore in what follows we will
assume $q>0,\omega>0$, and the results for the other regimes can be extrapolated from the above
symmetries.

There are two regimes which we
will study separately. One is $\omega > 2J\sin\frac{q}{2}$ where ${\rm Im}[\Pi^R]=\Pi^K=0$,
and the other is $\omega < 2J\sin\frac{q}{2}$ where ${\rm Im}[\Pi^R]\neq0, \Pi^K\neq 0$.

\subsection{Evaluation for $\omega > 2J\sin\frac{q}{2}$}

In this regime, the integrand contains no poles, and the result is purely real:
$\mbox{Re}\left[\Pi^{R}(q,\omega)\right] = \Pi^{R}(q,\omega)\,$ and $\mbox{Im}\left[\Pi^{R}(q,\omega)\right] = 0$. One may safely take $\delta\rightarrow 0\,$ to find
\begin{eqnarray}
& &\Pi^{R}(q,\omega) =  \frac{-\cos\frac{q}{4}}{2\pi i\sqrt{\omega^2-(2J\sin\frac{q}{2})^{2}}}\nonumber\\
&\times& \left\{z_{+}\ln\left[\frac{1+\frac{\sin\frac{q}{4}}{z_{+}}}{1-\frac{\sin\frac{q}{4}}{z_{+}}}\right]-z_{-}\ln\left[\frac{1+\frac{\sin\frac{q}{4}}{z_{-}}}{1-\frac{\sin\frac{q}{4}}{z_{-}}}\right]\right\}\nonumber\\
& - & \frac{\sin\frac{q}{4}}{2\pi i\sqrt{\omega^2-(2J\sin\frac{q}{2})^{2}}}\nonumber\\
& \times& \left\{z_{+}\ln\left[\frac{1+\frac{\cos\frac{q}{4}}{z_{+}}}{1-\frac{\cos\frac{q}{4}}{z_{+}}}\right]-z_{-}\ln\left[\frac{1+\frac{\cos\frac{q}{4}}{z_{-}}}{1-\frac{\cos\frac{q}{4}}{z_{-}}}\right]\right\},
\label{eq:pir1}
\end{eqnarray}
with
\begin{eqnarray}
z_{\pm}^{2} &=& \frac{1}{2}\pm\frac{1}{2}\sqrt{1-\left(\frac{\omega}{2J\sin\frac{q}{2}}\right)^{2}}
\,\,\forall \omega < 2J\sin\frac{q}{2}\nonumber \\
&=& \frac{1}{2}\pm\frac{i}{2}\sqrt{\left(\frac{\omega}{2J\sin\frac{q}{2}}\right)^{2}-1}
\,\, \forall \omega > 2J\sin\frac{q}{2}
\label{zpm}
\end{eqnarray}
Note that our convention is to place the branch-cut of the logarithm on the negative real axis.
In this same regime of $\omega > 2J \sin\frac{q}{2}$
there are no roots to the argument of the delta function in the Keldysh component,
and we have $\Pi^{K}(q,\omega) = 0$. With $\Pi^{R}(q,\omega)\,$ purely real,
this regime lies outside the particle-hole continuum. In sub-section~\ref{mode1} we will demonstrate the
existence of an undamped collective mode lying just above the particle-hole continuum for repulsive
interactions only.

\subsection{Evaluation for $\omega < 2J\sin\frac{q}{2}$}

For $\omega < 2J\sin\frac{q}{2}\,$ the integrand generically contains poles.
We extract the real and imaginary parts in the usual way by writing,
\begin{eqnarray}
& & \int \frac{dk}{2\pi} \frac{f(k)}{\omega + i\delta + \epsilon_{k}-\epsilon_{k+q}} =
\int \frac{dk}{2\pi}{\cal P}\left(\frac{f(k)}{\omega + \epsilon_{k}-\epsilon_{k+q}}\right)\nonumber\\
& & -  i\pi\int \frac{dk}{2\pi} f(k) \delta(\omega +\epsilon_{k}-\epsilon_{k+q})\\
& \equiv & \mbox{Re}\left[\Pi^{R}\right] + i\mbox{Im}\left[\Pi^{R}\right]
\end{eqnarray}
where ${\cal P}$ denotes taking the principal value of the integral. We obtain
\begin{eqnarray}
& & \mbox{Re}\left[\Pi^{R}(q,\omega)\right] = \frac{-\cos\frac{q}{4}}{2\pi\sqrt{(2J\sin\frac{q}{2})^{2}-\omega^{2}}}\nonumber\\
&\times &\left\{z_{+}\ln\left|\frac{1+\frac{\sin\frac{q}{4}}{z_{+}}}{1-\frac{\sin\frac{q}{4}}{z_{+}}}\right|- z_{-}\ln\left|\frac{1+\frac{\sin\frac{q}{4}}{z_{-}}}{1-\frac{\sin\frac{q}{4}}{z_{-}}}\right|\right\}\nonumber\\
&-&\frac{\sin\frac{q}{4}}{2\pi\sqrt{(2J\sin\frac{q}{2})^{2}-\omega^{2}}}\nonumber\\
&\times &\left\{z_{+}\ln\left|\frac{1+\frac{\cos\frac{q}{4}}{z_{+}}}{1-\frac{\cos\frac{q}{4}}{z_{+}}}\right|-z_{-}\ln\left|\frac{1+\frac{\cos\frac{q}{4}}{z_{-}}}{1-\frac{\cos\frac{q}{4}}{z_{-}}}\right|\right\},
\end{eqnarray}
with $z_{\pm}\,$ defined in Eq.~(\ref{zpm}).
The results for the imaginary part, $\mbox{Im}\left[\Pi^{R}\right] = \frac{\Pi^{R}-\Pi^{A}}{2i}$,
and the Keldysh component subdivide the
particle-hole continuum into two sub-regions, $0 < \omega < 2J\sin^{2}\frac{q}{2}\,$
and $2J\sin^{2}\frac{q}{2} < \omega < 2J\sin\frac{q}{2}$.

\begin{figure}
\includegraphics[totalheight=6cm,]{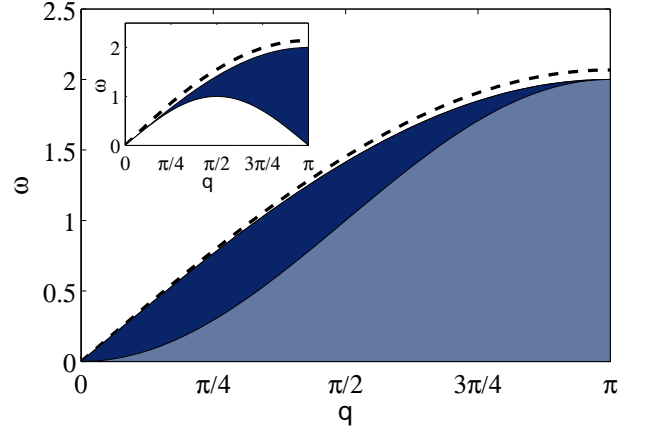}
\caption{Undamped mode (dashed line) above the extended particle-hole continuum
(shaded region). Inset: equilibrium continuum
and undamped mode.}
\label{plasmon1}
\end{figure}

For $0 < \omega < 2J\sin^{2}\frac{q}{2}\,$ one finds
\begin{eqnarray}
\mbox{Im}\left[\Pi^{R}(q,\omega)\right] & = & \frac{-1}{2\sqrt{(2J\sin\frac{q}{2})^{2}-\omega^{2}}}\left[\left(\cos\frac{q}{4}+\sin\frac{q}{4}\right)\right.\nonumber\\
& \times & \left.\sin\left[\frac{1}{2}\sin^{-1}\frac{\omega}{2J\sin\frac{q}{2}}\right]\right]\\
\Pi^{K}(q,\omega) & = & \frac{i}{2\sqrt{(2J\sin\frac{q}{2})^{2}-\omega^{2}}}\nonumber\\
&\times & \left[\cos\left(\sin^{-1}\left(\frac{\omega}{2J\sin\frac{q}{2}}\right)\right) - 2\right].
\end{eqnarray}
whereas in the region $2J\sin^{2}\frac{q}{2} < \omega < 2J\sin\frac{q}{2}$, the result is
\begin{eqnarray}
\mbox{Im}\left[\Pi^{R}(q,\omega)\right]  & = & \frac{-1}{2\sqrt{(2J\sin\frac{q}{2})^{2}-\omega^{2}}}\nonumber\\
& \times & \left[\sin\frac{q}{4}\left(\sin\left[\frac{1}{2}\sin^{-1}\frac{\omega}{2J\sin\frac{q}{2}}\right]\right.\right.\nonumber\\
& + & \left.\left.\cos\left[\frac{1}{2}\sin^{-1}\frac{\omega}{2J\sin\frac{q}{2}}\right]\right)\right]\\
\Pi^{K}(q,\omega) & = & \frac{i}{2\sqrt{(2J\sin\frac{q}{2})^{2}-\omega^{2}}}\left[\cos\frac{q}{2} - 2\right].
\end{eqnarray}
The particle-hole continuum, which is the region in $\omega,q$ space where ${\rm Im}\Pi^R\neq 0$
is indicated as the shaded region in Fig.~\ref{plasmon1} and compared with the equilibrium (no quench)
result (inset).  The two different shadings refer to the
discontinuity in the functional forms of ${\rm Im}[\Pi]^R,\Pi^K$ across $\omega = 2 J \sin^2\frac{q}{2}$.
The consequences of these results are discussed below.

\subsection{Undamped mode for $\omega > 2J\sin\frac{q}{2}$ and $V_0 > 0$}\label{mode1}

The undamped mode is obtained from the solution of
\begin{eqnarray}
1-V_0\Pi^R(q,\omega) = 0\label{root}
\end{eqnarray}
with $\Pi^R(q,\omega)$ given in Eq.~(\ref{eq:pir1}).
For sufficiently small $V_0$, we need to identify the points where $\Pi^R$ diverges.
This occurs for $\omega = 2J \sin(q/2)$.
Thus in the limit of $\omega \rightarrow 2J\sin\frac{q}{2}$,
we can write $z_{\pm} = \sqrt{\frac{1}{2}}\left(1\pm\frac{i\epsilon}{4J\sin\frac{q}{2}}\right)\,$
with $\epsilon = \sqrt{\omega^{2}-\left(2J\sin\frac{q}{2}\right)^{2}}$.
The dominant contribution assuming that
$q\rightarrow 0$ (so that $\cos\frac{q}{4}\approx 1$) is given by
\begin{eqnarray}
\Pi^{R}(q,\omega) & \simeq & - \frac{\sin\frac{q}{4}}{2\pi i\sqrt{\omega^{2}-\left(2J\sin\frac{q}{2}\right)^{2}}}\nonumber\\
& \times & \left\{z_{+}
\ln\left[\frac{1+\sqrt{2} + i\frac{\epsilon\sqrt{2}}{4J\sin\frac{q}{2}}}
{1-\sqrt{2} + i\frac{\epsilon\sqrt{2}}{4J\sin\frac{q}{2}}}\right]\right.\nonumber\\
& - &\left.z_{-}\ln\left[\frac{1+\sqrt{2} - i\frac{\epsilon\sqrt{2}}{4J\sin\frac{q}{2}}}{1-\sqrt{2} - i\frac{\epsilon\sqrt{2}}{4J\sin\frac{q}{2}}}\right]\right\}.
\end{eqnarray}
Due to the branch-cut in the logarithm (chosen to be on the negative real axis),
in the limit $\epsilon \rightarrow 0$, the above expression can be further simplified to give
\begin{eqnarray}
\Pi^{R}(q,\omega) & \simeq & \frac{\sin\frac{q}{4}}{\sqrt{2}\sqrt{\omega^{2}-(2J\sin\frac{q}{2})^{2}}},
\end{eqnarray}
Thus from Eq.~(\ref{root})
we find a single undamped mode {\sl provided} $V_{0}\,$ is positive with a dispersion
\begin{equation}
\omega_{q} \simeq \theta(V_{0})J|q|\sqrt{1 + \frac{V_{0}^{2}}{32J^{2}}} \label{mod1}.
\end{equation}
This can be compared to the undamped mode in the equilibrium problem,\cite{dassarma2,dassarma}
\begin{equation}
\omega_{q}^{\mbox{\scriptsize eq}} \simeq J|q|\sqrt{1+\frac{V_{0}}{\pi J}},
\end{equation}
which exists for both attractive and repulsive interactions.
Thus, for repulsive interactions, the obtained
sound wave is qualitatively similar to the equilibrium case, but with a slightly
modified velocity of propagation, whereas for attractive interactions, no undamped modes
exist.

\subsection{Enhanced particle-hole continuum and effective temperature}

The highly broadened initial fermion distribution gives one way to define an effective temperature
in this non-equilibrium state (cf. Eq.~(\ref{eq:teffc})). By analogy with the equilibrium properties
of the particle-hole bubbles, one may define an effective temperature in terms of
the collective degrees of freedom. Since the system is out of equilibrium, this temperature will
in general depend on $\omega,q$ and \emph{also} the chosen correlation function
\begin{equation}
\frac{\Pi^{R}(q,\omega)-\Pi^{A}(q,\omega)}{\Pi^{K}(q,\omega)} = \tanh\left(\frac{\omega}
{2T^{\prime}_{\rm eff}(q,\omega)}\right).
\end{equation}
For small frequencies, this ratio yields
\begin{equation}
\frac{\Pi^{R}-\Pi^{A}}{\Pi^{K}} \xrightarrow{\omega\rightarrow 0} \frac{\omega}{2J\sin\frac{q}{2}}
\left(\sin\frac{q}{4}+\cos\frac{q}{4}\right),
\end{equation}
so that for $\omega \rightarrow 0$, we obtain an effective-temperature
\begin{equation}
T^{\prime}_{\mbox{\scriptsize eff}}(q,\omega\rightarrow 0)
= \frac{J\sin\frac{q}{2}}{\cos\frac{q}{4} + \sin\frac{q}{4}}.
\end{equation}

We argue below that $T^{\prime}_{\rm eff}(q,\omega\rightarrow 0)$ is responsible for smearing out the
particle-hole continuum in much the same way that temperature
in an equilibrium system does.
To see this recall that the particle-hole continuum
represents the region in the $(q,\omega)$-plane where a collective mode of frequency
$\omega\,$ and wave-vector $q\,$ is unstable to decay into single particle-hole excitations.
The upper and lower continuum boundaries in equilibrium at zero temperature are given
by~\cite{Pearson62,todani}
\begin{eqnarray}
\omega_{L}(q) = J\sin q,\\
\omega_{U}(q) = 2J\sin\frac{q}{2}.
\end{eqnarray}

A simple argument makes these boundaries plausible: consider the energy of a single
particle-hole excitation which is created by removing a particle of momentum $k\,$
and creating a particle of momentum $k+q$,
\begin{equation}
\omega(k,q) = \epsilon_{k+q}-\epsilon_{k} = 2J\sin\frac{q}{2}\sin\left(k+\frac{q}{2}\right).
\end{equation}
This excitation energy depends not only on the momentum of the excitation, but also on the
momentum of the original particle, $k$. For a half-filled band at zero temperature,
the occupation of fermions is $\left\langle c_{k}^{\dagger}c_{k}\right\rangle_{\mbox{\scriptsize eq}}
= \theta\left(\frac{\pi}{2}-|k|\right)$. The only momenta available for hole creation are
those with $|k|<k_{F} = \frac{\pi}{2}$. Because the cosine dispersion has maximal slope at
$k = \frac{\pi}{2}$, the maximum excitation energy occurs for a given $q\,$
with $k = \frac{\pi}{2}-\frac{q}{2}$. The smallest excitation energy for a given $q\,$
at zero temperature occurs at $k = \frac{\pi}{2}\,$ or $k = \frac{\pi}{2}-q$. Thus
\begin{eqnarray}
\omega_{\mbox{\scriptsize max}}(q) & = & \omega\left(\frac{\pi}{2}-\frac{q}{2},q\right) = 2J\sin\frac{q}{2},\\
\omega_{\mbox{\scriptsize min}}(q) & = & \omega\left(\frac{\pi}{2},q\right) = J\sin q.
\end{eqnarray}
which are just the upper and lower boundaries of the particle-hole continuum (inset Fig.~\ref{plasmon1}).
Now, consider
lowering $k_{F}$. Excitations of smaller energy for a given $q\,$ are now possible, and in the
limit $k_{F}\rightarrow 0$, we have $\omega_{\mbox{\scriptsize min}}\rightarrow 2J\sin^{2}\frac{q}{2}$.
The result is the same if one considers the opposite limit of $k_{F}\rightarrow \pi\,$
at zero temperature.

In the present non-equilibrium situation, we find a particle-hole continuum
($\mbox{Im}\left[\Pi^{R}(q,\omega)\right]\neq 0$) that extends below this lower-bound all the
way to $\omega=0$.
A finite temperature is known to smear out this lower boundary,\cite{derzhko2}
due to the smoothing out of the zero-temperature step function for the occupation probability.
It is interesting to note that the expressions for $\mbox{Im}\left[\Pi^{R}(q,\omega)\right]\,$
and $\Pi^{K}(q,\omega)\,$ are actually continuous across the line $\omega = 2J\sin^{2}\frac{q}{2}$,
with discontinuities appearing in their derivatives. In Fig.~\ref{plasmon1} we plot the undamped
collective mode dispersion with the particle-hole continuum represented by the shaded region.
The two different shadings are separated by the line $\omega = 2 J\sin^2\frac{q}{2}$.
The analogous plot for the equilibrium situation is shown in the inset.

\section{RPA for current carrying state} \label{quench2}

We now apply RPA to study the current carrying non-equilibrium steady state described in section~\ref{IIb}.
The Keldysh component of the fermion Green's function is,
\begin{eqnarray}
iG_{f}^{K}(k;t,t') & = & \left[\cos\theta_{k}(1-n_{k}-n_{-k}) \right.\nonumber\\
& - & \left.(n_{k}-n_{-k})\right]e^{-i\epsilon_{k}(t-t')}\label{Gkcurr}.
\end{eqnarray}
while the retarded Green's function is given in Eq.~(\ref{Gr}),
and $n_{k} = \theta(k)\theta(k_{0}-k)$. Eq.~(\ref{Gkcurr}) implies that the
distribution function for the  Jordan-Wigner fermions
in the current carrying post-quench state is not only broad as
for the zero current case, but is also asymmetric in $k$, with sharp discontinuities superimposed on it
(see Fig.~\ref{dist}).
Thus we will find that as for the zero-current case, the particle-hole
continuum here too is broadened (extending everywhere below the line $\omega_{\rm max}= 2J\sin\frac{q}{2}$), while
the sharp structure in the distribution gives rise to some discontinuities in the
expression for ${\rm Im}[\Pi^{R}]$, and the appearance of additional damped modes.

The particle-hole bubbles are now given by
\begin{eqnarray}
\Pi^{R}(q,\omega) & = & -\frac{1}{2}\int\frac{dk}{2\pi}\left[\frac{\cos\theta_{k}(1-n_{k}-n_{-k})}{\omega + i\delta - 2J\sin\frac{q}{2}\sin\left(k+\frac{q}{2}\right)}\right.\nonumber\\
&-& \frac{(n_{k}-n_{-k})}{\omega + i\delta - 2J\sin\frac{q}{2}\sin\left(k+\frac{q}{2}\right)}\nonumber\\
&-&\frac{\cos\theta_{k+q}(1-n_{k+q}-n_{-k-q})}{\omega + i\delta - 2J\sin\frac{q}{2}\sin\left(k+\frac{q}{2}\right)}\nonumber\\
&+&\left.\frac{(n_{k+q}-n_{-k-q})}{\omega + i\delta - 2J\sin\frac{q}{2}\sin\left(k+\frac{q}{2}\right)}\right]
\label{eq:pircur}
\end{eqnarray}
\begin{eqnarray}
\Pi^{K}(q,\omega)  =  \frac{i}{2}(2\pi)\int\frac{dk}{2\pi}\delta(\omega + \epsilon_{k}-\epsilon_{k+q})\nonumber\\
\times\left\{ \left[\cos\theta_{k}(1-n_{k}-n_{-k}) - (n_{k}-n_{-k})\right]\right.\nonumber\\
\times\left[\cos\theta_{k+q}(1-n_{k+q}-n_{-k-q}) - (n_{k+q}-n_{-k-q})\right]\nonumber\\
\left. -1\right\}.
\label{eq:pikcur}
\end{eqnarray}
As before two regions appear, one where $\omega > 2J\sin\frac{q}{2}$ for which
${\rm Im}[\Pi^R]=\Pi^K=0$, and the second being $\omega < 2J\sin\frac{q}{2}$
where a particle-hole continuum is found to exist. We discuss these two regions separately.

\subsection{Evaluation for $\omega > 2J\sin\frac{q}{2}$}

In this regime, as before, the result is entirely real and we let $\delta\rightarrow 0$.
We find it convenient to write $\Pi^{R} = \Pi^{(1)}+\Pi^{(2)}$, where $\Pi^{(1)}\,$
depends on the Bogoliubov angle, $\cos\theta_{k} = \left|\sin\frac{k}{2}\right|$, while
$\Pi^{(2)}$ contains the rest. As before it is convenient to summarize the symmetries
of the polarization bubbles. We find $\Pi^{(1)}(-q,\omega)=\Pi^{(1)}(q,\omega)$,
however due to current flow,
$\Pi^{(2)}(-q,\omega)= -\Pi^{(2)}(q,\omega)$. Similarly, ${\rm Re}[\Pi^{(1)}](q,-\omega)={\rm Re}
[\Pi^{(1)}](q,\omega)$, while ${\rm Re}[\Pi^{(2)}](q,-\omega)=-{\rm Re}
[\Pi^{(2)}](q,\omega)$. In the discussion that follows, we take $q>0,\omega>0$.

We find,
\begin{eqnarray}
& &\Pi^{(1)}(q,\omega)  = -\frac{\cos\frac{q}{4}}{4\pi i\sqrt{\omega^2-(2J\sin\frac{q}{2})^{2}}}\nonumber\\
& \times &\left\{\mathcal{F}\left(\sin\left[\frac{k_{0}}{2}+\frac{q}{4}\right]\right)+\mathcal{F}\left(\sin\left[-\frac{k_{0}}{2}+\frac{q}{4}\right]\right)\right\}\nonumber\\
&-&  \frac{\sin\frac{q}{4}}{4\pi i \sqrt{\omega^2 - (2J\sin\frac{q}{2})^{2}}}\nonumber\\
&\times&\left\{\mathcal{F}\left(\cos\left[\frac{k_{0}}{2}+\frac{q}{4}\right]\right) + \mathcal{F}\left(\cos\left[-\frac{k_{0}}{2}+\frac{q}{4}\right]\right)\right\},
\end{eqnarray}
where
\begin{equation}
\mathcal{F}(z) = z_{+}\ln\left[\frac{\left(1+\frac{z}{z_{+}}\right)}{\left(1-\frac{z}{z_{+}}\right)}\right] - z_{-}\ln\left[\frac{\left(1+\frac{z}{z_{-}}\right)}{\left(1-\frac{z}{z_{-}}\right)}\right].
\end{equation}

In the limit $k_{0}\rightarrow 0$, we recover the results of section~\ref{quench1}.
The remaining terms can be collected as
\begin{eqnarray}
& &\Pi^{(2)}  =  \frac{1}{4\pi}\left\{\int_{k_{0}-q/2}^{k_{0}+q/2}-\int_{-q/2}^{q/2}\right\}\nonumber\\
& \times & dk\left\{\frac{1}{\omega-2J\sin\frac{q}{2}\sin k} +\frac{1}{\omega + 2J\sin\frac{q}{2}\sin k}\right\}\\
& = & -\frac{1}{4\pi i \sqrt{\omega^2 - (2J\sin\frac{q}{2})^{2}}}\nonumber\\
& \times & \left\{\ln\left[\frac{1+\frac{\omega\tan\left(\frac{k_{0}}{2}+\frac{q}{4}\right)}{4J\sin\frac{q}{2}z_{+}^{2}}}{1-\frac{\omega\tan\left(\frac{k_{0}}{2}+\frac{q}{4}\right)}{4J\sin\frac{q}{2}z_{+}^{2}}}\right] -  \ln\left[\frac{1+\frac{\omega\tan\left(\frac{k_{0}}{2}-\frac{q}{4}\right)}{4J\sin\frac{q}{2}z_{+}^{2}}}{1-\frac{\omega\tan\left(\frac{k_{0}}{2}-\frac{q}{4}\right)}{4J\sin\frac{q}{2}z_{+}^{2}}}\right]\right.\nonumber\\
&-&  \ln\left[\frac{1+\frac{\omega\tan\left(\frac{k_{0}}{2}+\frac{q}{4}\right)}{4J\sin\frac{q}{2}z_{-}^{2}}}{1-\frac{\omega\tan\left(\frac{k_{0}}{2}+\frac{q}{4}\right)}{4J\sin\frac{q}{2}z_{-}^{2}}}\right]+  \ln\left[\frac{1+\frac{\omega\tan\left(\frac{k_{0}}{2}-\frac{q}{4}\right)}{4J\sin\frac{q}{2}z_{-}^{2}}}{1-\frac{\omega\tan\left(\frac{k_{0}}{2}-\frac{q}{4}\right)}{4J\sin\frac{q}{2}z_{-}^{2}}}\right]\nonumber\\
&-&\left.  2\ln\left[\frac{1+\frac{\omega\tan\left(\frac{q}{4}\right)}{4J\sin\frac{q}{2}z_{+}^{2}}}{1-\frac{\omega\tan\left(\frac{q}{4}\right)}{4J\sin\frac{q}{2}z_{+}^{2}}}\right]+  2\ln\left[\frac{1+\frac{\omega\tan\left(\frac{q}{4}\right)}{4J\sin\frac{q}{2}z_{-}^{2}}}{1-\frac{\omega\tan\left(\frac{q}{4}\right)}{4J\sin\frac{q}{2}z_{-}^{2}}}\right]\right\},
\end{eqnarray}

The consequence of the above expressions for $\Pi^{(1,2)}$ will be discussed in section~\ref{currundamped}.

\subsection{Evaluation for $\omega < 2J\sin\frac{q}{2}$}

In this regime, the real part of $\Pi^{R}(q,\omega)\,$ is given by the principal value of the integral
in Eq.~(\ref{eq:pircur}).
Writing
$\mbox{Re}\left[\Pi^{R}(q,\omega)\right] = {\rm Re}[{\Pi}^{(1)}] + {\rm Re}[{\Pi}^{(2)}]$, where
\begin{eqnarray}
& & {\rm Re}[{\Pi}^{(1)}](q,\omega)  =  -\frac{\cos\frac{q}{4}}{4\pi\sqrt{(2J\sin\frac{q}{2})^{2}-\omega^{2}}}\nonumber\\
&\times &\left\{\tilde{\mathcal{F}}\left(\sin\left[\frac{k_{0}}{2}+\frac{q}{4}\right]\right)+\tilde{\mathcal{F}}\left(\sin\left[-\frac{k_{0}}{2}+\frac{q}{4}\right]\right)\right\}\nonumber\\
& - & \frac{\sin\frac{q}{4}}{4\pi\sqrt{(2J\sin\frac{q}{2})^{2}-\omega^{2}}}\nonumber\\
& \times & \left\{\tilde{\mathcal{F}}\left(\cos\left[\frac{k_{0}}{2}+\frac{q}{4}\right]\right) + \tilde{\mathcal{F}}\left(\cos\left[-\frac{k_{0}}{2}+\frac{q}{4}\right]\right)\right\},
\label{eq:repircur1}
\end{eqnarray}
where
\begin{equation}
\tilde{\mathcal{F}}(z) = z_{+}\ln\left|\frac{\left(1+\frac{z}{z_{+}}\right)}{\left(1-\frac{z}{z_{+}}\right)}\right| - z_{-}\ln\left|\frac{\left(1+\frac{z}{z_{-}}\right)}{\left(1-\frac{z}{z_{-}}\right)}\right|,
\end{equation}
and
\begin{eqnarray}
{\rm Re}[\Pi^{(2)}](q,\omega)  =  -\frac{1}{4\pi\sqrt{(2J\sin\frac{q}{2})^{2}-\omega^{2}}}\nonumber\\
\times  \left\{\ln\left|\frac{1+\frac{\omega\tan\left(\frac{k_{0}}{2}+\frac{q}{4}\right)}{4J\sin\frac{q}{2}z_{+}^{2}}}{1-\frac{\omega\tan\left(\frac{k_{0}}{2}+\frac{q}{4}\right)}{4J\sin\frac{q}{2}z_{+}^{2}}}\right| -  \ln\left|\frac{1+\frac{\omega\tan\left(\frac{k_{0}}{2}-\frac{q}{4}\right)}{4J\sin\frac{q}{2}z_{+}^{2}}}{1-\frac{\omega\tan\left(\frac{k_{0}}{2}-\frac{q}{4}\right)}{4J\sin\frac{q}{2}z_{+}^{2}}}\right|\right.\nonumber\\
  -  \ln\left|\frac{1+\frac{\omega\tan\left(\frac{k_{0}}{2}+\frac{q}{4}\right)}{4J\sin\frac{q}{2}z_{-}^{2}}}{1-\frac{\omega\tan\left(\frac{k_{0}}{2}+\frac{q}{4}\right)}{4J\sin\frac{q}{2}z_{-}^{2}}}\right|
 +  \ln\left|\frac{1+\frac{\omega\tan\left(\frac{k_{0}}{2}-\frac{q}{4}\right)}{4J\sin\frac{q}{2}z_{-}^{2}}}{1-\frac{\omega\tan\left(\frac{k_{0}}{2}-\frac{q}{4}\right)}{4J\sin\frac{q}{2}z_{-}^{2}}}\right|\nonumber\\
 -  \left.2\ln\left|\frac{1+\frac{\omega\tan\left(\frac{q}{4}\right)}{4J\sin\frac{q}{2}z_{+}^{2}}}{1-\frac{\omega\tan\left(\frac{q}{4}\right)}{4J\sin\frac{q}{2}z_{+}^{2}}}\right|
 +  2\ln\left|\frac{1+\frac{\omega\tan\left(\frac{q}{4}\right)}{4J\sin\frac{q}{2}z_{-}^{2}}}{1-\frac{\omega\tan\left(\frac{q}{4}\right)}{4J\sin\frac{q}{2}z_{-}^{2}}}\right|\right\},
\label{eq:repircur2}
\end{eqnarray}
We do not give expressions for $\mbox{Im}\left[\Pi^{R}\right], \Pi^K$ as the boundaries
of the particle-hole continuum are the same as in Fig.~\ref{plasmon1}, though there are additional
discontinuities within the continuum besides the one along $\omega = 2 J \sin^2\frac{q}{2}$.
Instead in the subsequent
sections, by studying the divergences in ${\rm Re }[\Pi^R]$ we will identify a single undamped mode
for repulsive interactions and $k_0 < \pi/2$, and several damped modes for both repulsive
and attractive interactions.

\subsection{Undamped mode for $\omega > 2J\sin\frac{q}{2}$ and $V_0 > 0$}\label{currundamped}

In this section we demonstrate that an undamped collective mode survives
provided the current is not too large. As before, define
$\epsilon = \sqrt{\omega^{2}-\left(2J\sin\frac{q}{2}\right)^{2}}$.
In the limit $q\rightarrow 0\,$ and for small $k_{0}$, the most divergent terms in
$\Pi^{R}(q,\omega)\,$ are
\begin{eqnarray}
\Pi^{R}(q,\omega) & \simeq & -\frac{\sin\frac{q}{4}}{4\pi\sqrt{\left(2J\sin\frac{q}{2}\right)^{2}-\omega^{2}}}\left\{\mathcal{F}\left[\cos\left(\frac{k_{0}}{2}\right)\right]\right.\nonumber\\
& + & \left.\mathcal{F}\left[\cos\left(-\frac{k_{0}}{2}\right)\right]\right\}\\
& \simeq & \frac{\sin\frac{q}{4}}{2\pi i\sqrt{2}\sqrt{\omega^{2}
-(2J\sin\frac{q}{2})^{2}}}\nonumber\\
& \times &\left\{\ln\left[1-\sqrt{2}\cos\frac{k_{0}}{2} + i\frac{\epsilon\sqrt{2}\cos\frac{k_{0}}{2}}{2J\sin\frac{q}{2}}\right] \right.\nonumber\\
& - & \left.\ln\left[1-\sqrt{2}\cos\frac{k_{0}}{2} - i\frac{\epsilon\sqrt{2}\cos\frac{k_{0}}{2}}{2J\sin\frac{q}{2}}\right]\right\}
\end{eqnarray}
The above expression shows that provided $k_{0}<\frac{\pi}{2}$, which corresponds to
the logarithms having a branch-cut, we obtain
\begin{eqnarray}
\Pi^R(q,\omega)& = & \frac{\sin\frac{q}{4}}{\sqrt{2}\sqrt{\omega^{2}-(2J\sin\frac{q}{2})^{2}}},
\end{eqnarray}
Thus by setting $1 - V_{0}\Pi^{R} = 0$, we recover the same dispersion as in the absence of current
\begin{equation}
\omega_{q} \simeq J|q|\sqrt{1 + \frac{V_{0}^{2}}{32J^{2}}}\theta(V_0)\theta(\pi/2-k_0)
\label{mod2}
\end{equation}
Thus, the undamped mode is unchanged for a current
which is below the threshold value of $k_0 < \pi/2$.
On the other hand, for currents larger than this value ($k_0 >\pi/2$) and
for $q \ll k_0$, no undamped modes exist.

\subsection{Damped modes for $\omega < 2J\sin\frac{q}{2}$}

In this regime, all modes are damped. We identify these damped modes by looking for solutions to
$1-V_{0}{\rm Re}[\Pi^{R}](q,\omega_{q})=0$. For $V_0\rightarrow 0$, all we need to do is identify
where $\mbox{Re}\left[\Pi^{R}\right]\rightarrow \pm \infty$. Then positive divergences
correspond to damped modes with repulsive interactions, while negative divergences correspond
to damped modes with attractive interactions.

Upon examining Eqns.~(\ref{eq:repircur1}) and ~(\ref{eq:repircur2}), we find logarithmic divergences
in $\mbox{Re}\left[\Pi^{R}(q,\omega)\right]\,$ along the characteristic lines (for $\omega,q >0$)
\begin{eqnarray}
\omega_{1}(q) & = & 2J\sin^{2}\frac{q}{2},\\
\omega_{2}(q) & = & 2J\sin\frac{q}{2}\sin\left(k_{0} + \frac{q}{2}\right),\\
\omega_{3}^{\pm}(q) & = &  \pm 2J\sin\frac{q}{2}\sin\left(\frac{q}{2}-k_0\right),
\end{eqnarray}
Note that
$\omega_{2,3}$ coincide with the characteristic lines in the equilibrium problem with an
arbitrary Fermi momentum  $k_{0}$.\cite{derzhko2}
In the equilibrium problem, these lines represent boundaries across which
$\mbox{Im}\left[\Pi^{R}\right]\,$ undergoes a jump discontinuity. This is also the case here,
though we will focus our attention on the behavior of the real part.

\begin{figure}
\includegraphics[totalheight=6cm,]{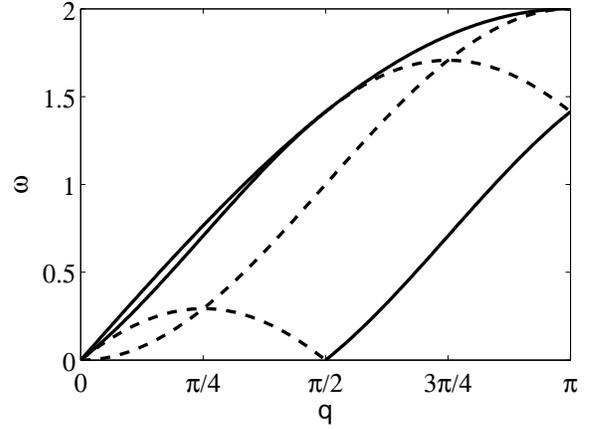}
\caption{Characteristic lines along which $\mbox{Re}\left[\Pi^{R}(q,\omega)\right]\,$
diverges, giving rise to damped modes for repulsive (solid lines) and attractive (dashed lines)
interactions for $k_{0} = \frac{\pi}{4}$. Only the top line $\omega = 2J\sin\frac{q}{2}\,$
corresponds to an undamped mode existing above the particle-hole continuum.}
\label{characteristics}
\end{figure}


One finds $\mbox{Re}\left[\Pi^{R}\right]\rightarrow +\infty\,$ along the line $\omega_{3}^{+}(q)\,$
for $2k_{0}<q < \pi\,$ and along the line $\omega_{2}(q)\,$ for $q < \pi - 2k_{0}$.
These correspond to damped collective modes for repulsive interactions.
Furthermore, $\mbox{Re}\left[\Pi^{R}\right]\rightarrow -\infty\,$ along the line $\omega_{1}(q)\,$
for all $q \in (0,\pi)$, along $\omega_{3}^{-}(q)\,$ for $q<2k_{0}$, and along $\omega_{2}(q)\,$ for $q>\pi-2k_{0}$.
These negative divergences represent collective modes created by attractive interactions.
We plot these characteristic lines in Fig.~\ref{characteristics} and indicate whether
the mode exists for attractive or repulsive interactions.

Such damped modes are usually considered physically uninteresting\cite{dassarma} compared to any undamped
excitations in the system, as the damping makes these modes experimentally unobservable.
The divergences in $\Pi^{R}(q,\omega)\,$ that give rise to these damped modes are of a different
nature than those giving rise to the undamped mode. To see this consider the case
of $\omega \simeq \omega_1 = 2J\sin\frac{q}{2}\left(\sin\frac{q}{2} \pm \epsilon\right)\,$
for small $\epsilon$. The dominant contribution is given by
\begin{equation}
\mbox{Re}\left[\Pi^{R}(q,\omega)\right] \simeq \frac{-1}{8\pi J\sin\frac{q}{2}\left|\cos\frac{q}{2}\right|}\left[2\log\frac{A_q}{\epsilon}\right].
\label{eq:pirch1}
\end{equation}
where $A_q$ is a $q$-dependent factor.

Solving $1=V_{0}\mbox{Re}\Pi^{R}\left[(q,\omega_{q})\right]\,$
to leading order in $\epsilon$, one finds
\begin{equation}
\mbox{Re}\left[\omega_{q}\right] \simeq 2J\sin\frac{q}{2}\left(\sin\frac{q}{2} \pm A_qe^{-(2\pi J\sin q)/|V_{0}|}\right),
\end{equation}
where $q>0\,$ is assumed. In general, for any of the characteristic lines described above, the
ansatz $\omega = \omega_{c} \pm 2J\epsilon\sin\frac{q}{2}\,$ leads to a divergence of the form
$\Pi^{R}(q,\omega) \sim \pm \log \frac{1}{\epsilon}$.
Each logarithmic divergence corresponds to two damped modes lying exponentially close to
each characteristic line.

One may study the modes near $\omega \approx \omega_{2}(q)\,$ in a way similar to our analysis for the
modes near $\omega_1$.
For $\omega\approx \omega_{2}(q)\pm 2J\epsilon\sin\frac{q}{2}$, one finds
\begin{equation}
\mbox{Re}\left[\Pi^{R}(q,\omega)\right] \simeq \frac{1+\sin\frac{k_{0}}{2}}
{8\pi J \sin\frac{q}{2}\left|\cos\left(k_{0}+\frac{q}{2}\right)\right|}\log\frac{A^{\prime}_q}{\epsilon},
\label{eq:pirch2}
\end{equation}
which gives rise to damped modes for repulsive interactions ($V_{0}>0$) when $q <\pi - 2 k_0$ with
\begin{eqnarray}
& & \mbox{Re}\left[\omega_{q}\right]\simeq 2J\sin\frac{q}{2}
\left(\sin\left(\frac{q}{2}+k_{0}\right)\right.\nonumber\\
& \pm &\left. A_q^{\prime}
\exp\left[-\frac{8\pi J\sin\frac{q}{2}\left|\cos\left(k_{0}+\frac{q}{2}\right)\right|}{V_{0}\left(1+\sin\frac{k_{0}}{2}\right)}\right]\right).
\end{eqnarray}
The results for the other characteristic line $\omega_3^+$ is similar and we do not discuss it further.

\section{Summary and Conclusions}
\label{conclu}

In this paper, we have applied the RPA to study the effect of weak Ising interactions
in a non-equilibrium steady state
of the $XXZ\,$ spin chain.
This non-equilibrium state
was created in two different ways. One is by quenching from the ground state of the
transverse-field Ising model at critical magnetic
field to the $XX$-model. The second was to modify the Hamiltonian before the quench
by adding
Dzyaloshinskii-Moriya interactions. This had the effect of creating a
current carrying state.

The RPA for both the steady-states shows the existence of
a single, undamped, collective mode for repulsive interactions which is qualitatively similar
to the sound mode in equilibrium, but with quantitative changes to the mode velocity
(c.f. Eq.~(\ref{mod1})). However if the current is larger than a threshold value, this undamped mode
ceases to exist in the long-wavelength limit (c.f. Eq.~(\ref{mod2})). The primary effect of the
quench is to give rise to a highly broadened distribution function (c.f. Fig.~\ref{dist},
Eq.~(\ref{dist2})) which results in an enhanced particle-hole continuum. The boundaries of the particle-hole
continuum are shown in Fig.~\ref{plasmon1}.
Thus for attractive interactions either no modes are found for the first steady-state, or some
damped collective modes are found for the steady-state with current.

These results, and in particular the generation of a finite friction due to an out of
equilibrium situation, are rather generic and
do not depend on the details of the non-equilibrium steady-state. Further, the upper boundary
of the particle-hole continuum occurs at $\omega_{\rm max} = 2 J\sin\frac{q}{2}$ and is related to the
fact that the system is on a lattice, and therefore the excitations have a maximum velocity.
If instead a quadratic dispersion for the fermions is adopted, then there is no
upper-limit to the velocity of excitations. This together with the fact
that immediately after a quench, the Fermi-distribution is very broad with no well defined
$k_F$ will further enhance the upper boundary of the particle-hole
continuum, damping even the mode with repulsive interactions.

An important future direction for research is to explore how these results change when
an explicit time-dependence on $J^z$ is introduced. In particular it is important to
understand how slowly $J^z$ has to be turned on in order to recover the
results of this paper.

{\sl Acknowledgments:} This work was supported by
NSF DMR (Grant No. 1004589) (JL and AM) and by the Swiss SNF under MaNEP and Division II
(TG).

\end{document}